\documentclass[aps,ams,amsmath,prx,longbibliography,twocolumn,superscriptaddress]{revtex4-1}
\usepackage{graphics}
\usepackage{graphicx}
\usepackage{epstopdf,epsfig}
\usepackage{amsmath}
\usepackage{amssymb}
\usepackage{amsfonts}
\usepackage{mathrsfs}
\usepackage{bm}
\usepackage{xcolor}
\usepackage{bbm}
\usepackage{hyperref}
\usepackage[normalem]{ulem}
\usepackage{comment}
\usepackage{soul}

\newcommand{\dd}{d}
\newcommand{\eps}{\varepsilon}
\newcommand{\kF}{k_F}
\newcommand{\aB}{a_B}
\newcommand{\A}{\mathcal{A}}

\begin{document}

\title{Fluctuation electrodynamics of quantum capacitance in electron bilayers}

\author{Dmitry Zverevich}
\affiliation{Department of Physics, University of Wisconsin-Madison, Madison, Wisconsin 53706, USA}

\author{Alberto F. Morpurgo}
\affiliation{Department of Quantum Matter Physics, University of Geneva, Quai Ernest-Ansermet 24, 1211 Geneva, Switzerland}
\affiliation{Department of Applied Physics, University of Geneva, 24 Quai Ernest Ansermet, Geneva, CH-1211 Switzerland}
\affiliation{Geneva Quantum Center, University of Geneva, Switzerland}

\author{Alex Levchenko}
\affiliation{Department of Physics, University of Wisconsin-Madison, Madison, Wisconsin 53706, USA}

\date{October 30, 2026}

\begin{abstract}
Capacitance provides a thermodynamic probe of interacting electrons. 
It characterizes the energy cost of transferring charge between conductors, 
which, in low-dimensional systems, is governed by exchange and correlation effects in addition to electrostatic contributions.
We develop a theory of the quantum capacitance of electron
double layers, semiconductor quantum wells as well as monolayer and bilayer
graphene devices, focusing on the contribution generated by interlayer
correlations. Within a functional-integral formulation we show that the
separation-dependent part of the ground-state energy is, at the level of ring
diagrams, exactly the nonretarded Lifshitz expression for the van der Waals
energy of two conducting sheets, with reflection amplitudes built from the
layer polarizabilities. The interlayer correction to the inverse capacitance
is the second density derivative of this energy: a Casimir compressibility.
The zero-point fluctuations that generate Casimir forces between mirrors are
here the coupled plasmons of the bilayer, and their contribution to the
capacitance is obtained in closed form; 
its sign shows that interlayer correlations
oppose charging at high density. A Gell-Mann-Brueckner analysis gives the
high-density limit in terms of universal functions. For
graphene we find that monolayers behave like the electron gas, while
bilayer graphene is anomalous: interband screening suppresses the correction
by orders of magnitude and, within the two-band model, can even reverse
its sign in the experimentally relevant
range of separations. We delineate the limits of the theory, identifying the dilute
Wigner-crystal regime and electron-hole double layers near exciton condensation
as regimes where the capacitance becomes a probe of interlayer pairing.
\end{abstract}

\maketitle

\section{Introduction}
\label{sec:intro}

Electron double layers, consisting of two conducting sheets separated by a nanometer- to micron-scale insulating barrier that suppresses interlayer tunneling 
while preserving Coulomb coupling, have long provided a versatile platform for studying electron correlations. 
Since the two layers are coupled exclusively through the Coulomb forces, any interlayer observable necessarily reflects interaction effects.
Coulomb drag, in which a current in one layer transfers momentum to the
other, is the transport realization of this idea~\cite{Gramila1991,
Narozhny2016}. Capacitance and penetration-field measurements are its
thermodynamic counterpart. Eisenstein \textit{et al.}
\cite{Eisenstein1992,Eisenstein1994} showed that the electric field leaking
through an imperfectly screening two-dimensional electron gas (2DEG)
measures the compressibility of that layer. Using a double quantum well,
they demonstrated that $d\mu/dn$ of a dilute 2DEG is negative, i.e., that
exchange and correlation overwhelm the kinetic energy, confirming a
prediction made a decade earlier~\cite{Bello1981,ChenskiiTkach1984}.

The graphene era turned these techniques into precision
tools. Quantum-capacitance and compressibility measurements on monolayer and
bilayer graphene~\cite{Xia2009,Ponomarenko2010,Henriksen2010,Yu2013,
Martin2008}, including the dual-gated geometries in which density and layer
polarization are tuned independently~\cite{Young2012,YoungLevitov2011},
established capacitance as a spectroscopic probe of band structure and
interactions on an equal footing with transport. A number of
discoveries have since been made with capacitance bridges: correlated composite-fermion phases in bilayer
graphene Landau levels~\cite{Zibrov2017}, the half- and quarter-metal
ferromagnets of rhombohedral trilayer graphene revealed by
penetration-field capacitance on graphite-gated van der Waals
heterostructures~\cite{Zhou2021}, and, most recently, flat-band surface-state
superconductivity in thick rhombohedral graphite detected by layer-resolved
capacitance~\cite{Guo2025}. On the theory side these experiments
revived interest in the capacitance of interacting double
layers~\cite{Luryi1988,KoppMannhart2009,SkinnerShklovskii2010,
SkinnerFogler2010,Profumo2010,Parhizgar2017,Berthod2021,Zhang2019},
with particular attention to the dilute regime where the electron liquid is
nearly a Wigner crystal and the capacitance can exceed its geometric
value~\cite{Bello1981,SkinnerShklovskii2010,LiAshoori2011}. The recent
observation of robust Wigner crystals and bilayer Wigner crystals in
transition-metal dichalcogenide heterostructures~\cite{Smolenski2021,
ZhouWC2021,LiCrommie2021,Zhou2026} shows that this strongly correlated
regime is now accessible experimentally.

An important part of this theory is, however, missing. The measured
inverse capacitance of a double layer separates into three parts: the
geometric term, the single-layer compressibilities (the conventional
``quantum capacitance''~\cite{Luryi1988}), and a remainder generated by
correlations between the layers. The single-layer parts are known
essentially exactly from quantum Monte Carlo~\cite{TanatarCeperley1989,
Attaccalite2002}. The interlayer part is the poorly understood one: it has
been computed numerically in specific models~\cite{Parhizgar2017}, estimated
classically in the Wigner-crystal regime~\cite{SkinnerShklovskii2010}, and
identified diagrammatically as a ``cross quantum capacitance''
\cite{Berthod2021,Zhang2019}. Its general structure, controlled
limits, asymptotic behavior, and sign have not been
established. This is the purpose of the present paper.

The organizing principle of our approach is that this problem belongs
to Casimir physics. The Casimir force between conductors~\cite{Casimir1948} and its
generalization to real materials, the Lifshitz theory of van der Waals
forces~\cite{Lifshitz1956,Dzyaloshinskii1961}, express the
separation-dependent part of the ground-state energy of two bodies through
their electromagnetic reflection amplitudes; the modern state of this field
is documented in several monographs~\cite{Bordag2009,Milton2001,
Mostepanenko1997,Dalvit2011} and reviews~\cite{Woods2016}. Casimir-force
experiments now operate at separations of $100$~nm--$2\,\mu$m with
sub-percent precision, enough to resolve how the fluctuating fields penetrate
real electrodes and to discriminate between models of their low-frequency
response~\cite{GarciaSanchez2012}. We show that the interlayer
correlation energy of an electron bilayer in the ring (random-phase)
approximation is the nonretarded Lifshitz energy of two two-dimensional
sheets. The reflection amplitude of each sheet is built from its density
response function. In the regime relevant to capacitors the zero-point
photon modes of the Casimir problem are replaced by the coupled plasma
oscillations of the two layers. The interlayer capacitance correction is the
second density derivative of their zero-point energy, a quantity we call the
Casimir compressibility. The energy part of this correspondence is
known in the theory of dispersion forces. Sernelius and Bj\"ork computed the van der
Waals attraction of two quantum wells~\cite{SerneliusBjork1998}; Bostr\"om,
Sernelius, and earlier Barash found its anomalous fractional power
laws~\cite{BostromSernelius2000,Barash1989}; Dobson, White, and Rubio
established the corresponding asymptotic benchmarks for layered and graphene
systems~\cite{Dobson2006,Sarabadani2011}. The capacitance has not
been considered in this context. It requires the dependence of the
reflection amplitudes on the layer densities, which is not probed by a
Casimir force measured at fixed density. Here we work out these consequences
quantitatively for realistic systems.
The same interlayer fluctuations, evaluated on the real frequency axis,
produce Coulomb drag~\cite{Narozhny2016}; the Casimir compressibility is the
equilibrium moment of the same fluctuation spectrum, so the present work
places drag and capacitance within one framework.

The paper is organized as follows. Section~\ref{sec:setup} defines the
model, the observables, and the separation of the inverse capacitance into
geometric, single-layer, and interlayer parts. Section~\ref{sec:functional}
derives the interlayer correlation energy from the fermionic functional
integral, emphasizing the matrix structure in layer space that distinguishes
the bilayer from the textbook single-layer analysis.
Section~\ref{sec:casimir} proves the equivalence to Lifshitz theory,
quantifies the neglect of retardation, discusses finite temperature and the
limits of the additive (ring) approximation, and formulates the Casimir
compressibility. Section~\ref{sec:2deg} evaluates the theory for the 2DEG
double layer: the exact high-density (Gell-Mann--Brueckner) limit, the ring
resummation, the plasmon zero-point asymptotics with their universal
constants, and order-of-magnitude estimates (Fig.~\ref{fig:2deg}).
Section~\ref{sec:graphene} presents the corresponding results for monolayer
and bilayer graphene (Figs.~\ref{fig:mlg} and \ref{fig:blg}), including the
anomalous sign structure of the bilayer case and the limits of validity near
the neutrality point. Section~\ref{sec:discussion} places the results
against the known limiting regimes, from the classical dipole regime of
dilute bilayers to exciton condensation in electron-hole double layers, and
identifies the open problems. Section~\ref{sec:summary} summarizes.
Technical material is collected in Appendixes~\ref{app:lindhard}--%
\ref{app:numerics}.

\section{Setup and observables}
\label{sec:setup}

\subsection{Model}
We consider two parallel two-dimensional electron layers separated by a
distance $d$ in a uniform dielectric environment $\bar\epsilon$ (Gaussian
units). Tunneling is absent; the layers are coupled only through the Coulomb
interaction. Each layer hosts a translationally invariant electron liquid
with areal density $n_\ell$ ($\ell=1,2$) on its own neutralizing background,
so that each layer is neutral in the reference state. The intra- and
interlayer interactions are
\begin{equation}
v(q)=\frac{2\pi e^2}{\bar\epsilon q},
\qquad
v_{12}(q)=\frac{2\pi e^2}{\bar\epsilon q}\,e^{-qd}.
\label{eq:vq}
\end{equation}
Three microscopic realizations are treated in parallel: (i) the parabolic
2DEG (mass $m$, spin degeneracy 2), for which we use effective atomic units
[$\hbar=m=e^2/\bar\epsilon=1$, lengths in Bohr radius $\aB=\bar\epsilon\hbar^2/me^2$,
energies in Hartree unit $\mathrm{Ha}=e^2/\bar\epsilon\aB$], with $n=\kF^2/2\pi$,
density of states $\nu_0=m/\pi\hbar^2$, interaction parameter
$r_s=(\pi n)^{-1/2}/\aB$, and Thomas-Fermi wave vector
$\kappa=2\pi e^2\nu_0/\bar\epsilon=2/\aB$; (ii) doped monolayer graphene
(MLG), a chirality-one Dirac liquid with velocity $v$, degeneracy $g=4$, and
coupling $\alpha_{\rm g}=e^2/\bar\epsilon\hbar v$; and (iii) doped Bernal
bilayer graphene (BLG) in the two-band approximation~\cite{McCann2006}, a
chirality-two liquid with mass $m$ and $g=4$. For graphene systems
$n=g\kF^2/4\pi$.

\subsection{Capacitance as an energy curvature}
\label{sec:thermo}
Let $\eps(n_1,n_2;d)$ denote the ground-state energy per unit area.
Transferring an areal density $\delta n$ between the plates at fixed total
density, $n_{1,2}=\bar n_{1,2}\pm\delta n$, requires the voltage
$eU=\partial\eps/\partial\delta n$, so the differential capacitance per unit
area is
\begin{equation}
\frac{1}{C}
=\frac{1}{e^2}\frac{\dd^2\eps}{\dd(\delta n)^2}
=\frac{1}{e^2}\left[
\frac{\partial^2\eps}{\partial n_1^2}
+\frac{\partial^2\eps}{\partial n_2^2}
-2\frac{\partial^2\eps}{\partial n_1\partial n_2}\right].
\label{eq:capdef}
\end{equation}
Equation~\eqref{eq:capdef} agrees with the curvature formula used in
Ref.~\cite{Parhizgar2017} and, in linear-response language, with the general
result of Ref.~\cite{Berthod2021}. The energy separates into three
distinct contributions,
\begin{equation}
\eps=\frac{2\pi e^2 d}{\bar\epsilon}\,\delta n^2
+\eps_0(n_1)+\eps_0(n_2)
+\eps_{\rm int}(n_1,n_2;d),
\label{eq:decomp}
\end{equation}
namely the electrostatic (Hartree) energy of the transferred charge, the
energies of the two isolated layers, and the interlayer correlation energy,
which contains all the $d$ dependence beyond electrostatics. Here
$\eps_0(n)$ is the ground-state energy per unit area of an isolated neutral
layer of density $n$ on its own neutralizing background, so that
$\mu_0(n)=\dd\eps_0/\dd n$ below is the standard chemical potential of the
two-dimensional electron liquid. The entire electrostatic energy of the
transferred charge is contained in the first term, which is exact for
uniform sheets; the exchange-correlation energy of a uniform layer does not
depend on where its neutralizing charge resides, so defining $\eps_0$
instead at a fixed background would differ from our convention only by that
Hartree term. Accordingly,
\begin{align}
&C^{-1}=\frac{4\pi d}{\bar\epsilon}
+\frac{1}{e^2}\left[\frac{\partial\mu_0}{\partial n}\Big|_{n_1}
+\frac{\partial\mu_0}{\partial n}\Big|_{n_2}\right]
+\delta C^{-1}_{\rm inter},
\label{eq:threeparts}\\
&\delta C^{-1}_{\rm inter}\equiv\frac{1}{e^2}\Bigl[\partial^2_{n_1}+\partial^2_{n_2}
-2\,\partial_{n_1}\partial_{n_2}\Bigr]\,\eps_{\rm int}.
\label{eq:dcint}
\end{align}
The first term of Eq.~\eqref{eq:threeparts} is the geometric capacitance;
the second contains the single-layer inverse compressibilities, negative at
low density~\cite{Bello1981,Eisenstein1992,Eisenstein1994} and known
essentially exactly from quantum Monte Carlo parametrizations
\cite{TanatarCeperley1989,Attaccalite2002,GoriGiorgi2004}. The third term is
the subject of this paper. The derivatives in Eqs.~\eqref{eq:threeparts}
and \eqref{eq:dcint} are evaluated at the actual layer densities
$n_{1,2}=\bar n\pm\delta n$ of the biased state; Eq.~\eqref{eq:threeparts}
is the differential capacitance at that working point. For a symmetric
device at zero bias both arguments equal $\bar n$, which is the case
evaluated in Secs.~\ref{sec:2deg} and \ref{sec:graphene}.

Two remarks about this separation are in order. First, Eq.~\eqref{eq:decomp}
defines $\eps_0(n)$ as the energy of an isolated layer: since the dielectric
environment is uniform, the intralayer interaction carries no dependence on
$d$, and every interlayer effect, including the modification of intralayer
screening by the presence of the second layer, resides, by construction, in
$\eps_{\rm int}$. This convention differs from analyses formulated in
terms of the polarizabilities of the coupled capacitor~\cite{Berthod2021}.
There the ``intralayer'' response $\Pi_{\ell\ell}$ is itself dressed by the
other plate, and only the off-diagonal $\Pi_{12}$ is counted as a cross
term. The two conventions are related by moving the diagonal
derivatives $\partial^2\eps_{\rm int}/\partial n_\ell^2$, the $d$-dependent
renormalization of each layer's compressibility, from one category to the
other. The quantities that do not depend on the convention are
$\eps_{\rm int}$ and its full Hessian, which is what we compute. As shown in Sec.~\ref{sec:capresult}, the
diagonal part carries about two thirds of the interlayer correction and the
pure cross part one third; identifying the interlayer effect with the cross
term alone would therefore understate it substantially.
Second, it is convenient to quote the interlayer correction as a shift of the
effective capacitor thickness $d^*\equiv\bar\epsilon/4\pi C$,
\begin{equation}
\delta d^*_{\rm inter}=\frac{\bar\epsilon}{4\pi}\,
\delta C^{-1}_{\rm inter}.
\label{eq:dstar}
\end{equation}
We note that the full Hessian $\partial\mu_i/\partial n_j$ also controls the
penetration-field capacitance and the Eisenstein ratio~\cite{Eisenstein1992,
YoungLevitov2011,Parhizgar2017}; measuring several capacitances of one
device therefore separates the mixed derivative
$\partial_{n_1}\partial_{n_2}\eps$ from the diagonal terms, a point we
return to in Sec.~\ref{sec:discussion}.

\section{Interlayer correlation energy from the functional integral}
\label{sec:functional}

\subsection{Partition function and layer-space structure}
We start from the imaginary-time coherent-state representation of the grand
partition function,
\begin{equation}
Z=\int D[\bar\psi,\psi]\,e^{-S[\bar\psi,\psi]},\qquad
S=S_0+S_{\rm int},
\label{eq:Z}
\end{equation}
with the free action
\begin{equation}
S_0=\int_0^\beta\!\dd\tau\sum_{\ell=1,2}\sum_{\bm k}
\bar\psi_{\ell\bm k}\bigl(\partial_\tau+\xi_{\ell\bm k}\bigr)\psi_{\ell\bm k},
\end{equation}
where $\xi_{\ell\bm k}$ is the band dispersion of layer $\ell$ measured from
its chemical potential (band and spin indices are implicit), and the
density-density interaction
\begin{subequations}\label{eq:Sint}
\begin{align}
S_{\rm int}=\frac{1}{2\A}\int_0^\beta\!\dd\tau\sum_{\bm q\neq0}
\sum_{\ell\ell'}\rho_{\ell,-\bm q}\,V_{\ell\ell'}(q)\,\rho_{\ell',\bm q},
\\
\hat V(q)=
\begin{pmatrix}
v & v_{12}\\ v_{12} & v
\end{pmatrix}.
\end{align}
\end{subequations}
Here $\rho_{\ell,\bm q}=\sum_{\bm k}\bar\psi_{\ell,\bm k}\psi_{\ell,\bm k+\bm q}$
and the $\bm q=0$ component is cancelled by the neutralizing backgrounds.
The only difference from the single-layer problem is that the interaction is
a $2\times2$ matrix in layer space, with the entire $d$ dependence residing
in the off-diagonal element $v_{12}(q)=v(q)e^{-qd}$ of Eq.~\eqref{eq:vq}.
Since the layers are distinguishable and tunneling is absent, there are no interlayer
Green's functions: interlayer exchange vanishes identically, and the
interlayer energy starts at second order in $v_{12}$ (it is pure
correlation). This removes the second-order exchange term that complicates
the original Gell-Mann-Brueckner analysis of the bulk electron
gas~\cite{GellMannBrueckner1957}. The matrix $\hat V(q)$ is
invertible for any $d>0$, $\det\hat V=v^2(1-e^{-2qd})$, so a
Hubbard-Stratonovich decoupling in the charge channel is well defined.

Introducing a two-component real bosonic field $\phi_\ell(\bm q,\tau)$
conjugate to the layer densities and decoupling Eq.~\eqref{eq:Sint},
\begin{equation}
Z=\int D\phi\;e^{-\frac{1}{2}\int\phi^{T}\hat V^{-1}\phi}
\int D[\bar\psi,\psi]\,
e^{-\sum_\ell\bar\psi_\ell(\hat G^{-1}_{0\ell}+i\phi_\ell)\psi_\ell},
\end{equation}
the fermions appear quadratically and can be integrated
out exactly:
\begin{equation}
Z=\int D\phi\;
\exp\Bigl\{-\tfrac12\!\int\!\phi^{T}\hat V^{-1}\phi
+\sum_\ell\mathrm{Tr}\ln\bigl[\hat G^{-1}_{0\ell}+i\phi_\ell\bigr]\Bigr\}.
\label{eq:Zphi}
\end{equation}
The $\mathrm{Tr}\ln$ is diagonal in the layer index (each fermion couples
only to its own field $\phi_\ell$) while the fields are coupled through
$\hat V^{-1}$. All interlayer physics is therefore mediated by the boson
propagator, i.e., by the screened Coulomb interaction; this is the formal
statement of the absence of interlayer exchange.

\subsection{One-loop free energy}
Expanding the $\mathrm{Tr}\ln$ in Eq.~\eqref{eq:Zphi} to quadratic order in
$\phi_\ell$ generates $\frac12\phi^{T}\hat\chi_0\phi$, with
$\hat\chi_0=\mathrm{diag}(\chi_{0,1},\chi_{0,2})$ the matrix of free
polarization bubbles, $\chi_{0,\ell}(q,i\Omega_m)<0$ on the Matsubara axis. The remaining Gaussian integral gives the
fluctuation grand potential per unit area
\begin{equation}
\frac{\Delta\Omega}{\A}
=\frac{1}{2\beta}\sum_{\Omega_m}\int\!\frac{\dd^2q}{(2\pi)^2}\,
\ln\det\bigl[\mathbbm{1}-\hat V(q)\hat\chi_0(q,i\Omega_m)\bigr]
+\;\cdots,
\label{eq:trlog}
\end{equation}
where the ellipsis denotes normal-ordering constants that are linear in
the layer densities $n_\ell$ and are therefore annihilated by the second
derivatives in Eq.~\eqref{eq:capdef}. Equation~\eqref{eq:trlog} is the bilayer
generalization of the familiar ring (RPA) exchange-correlation energy; the
only difference is the determinant over the layer space,
\begin{equation}
\det\bigl[\mathbbm{1}-\hat V\hat\chi_0\bigr]
=(1-v\chi_{0,1})(1-v\chi_{0,2})-v_{12}^2\,\chi_{0,1}\chi_{0,2}.
\label{eq:det}
\end{equation}
At $T\to0$, $\beta^{-1}\sum_{\Omega_m}\to
\int_{-\infty}^{\infty}\dd\xi/2\pi$. Subtracting the isolated-layer
logarithms, which renormalize $\eps_0(n_\ell)$ in Eq.~\eqref{eq:decomp}, the
interlayer part (the only $d$-dependent part) takes the compact form
\begin{align}
\frac{E_{\rm int}}{\A}
&=\frac{\hbar}{2}\int\!\frac{\dd^2q}{(2\pi)^2}
\int_{-\infty}^{\infty}\!\frac{\dd\xi}{2\pi}\,
\ln\left[1-\frac{v_{12}^2\,\chi_{0,1}\chi_{0,2}}
{(1-v\chi_{0,1})(1-v\chi_{0,2})}\right]
\nonumber\\
&=\frac{\hbar}{2}\int\!\frac{\dd^2q}{(2\pi)^2}
\int_{-\infty}^{\infty}\!\frac{\dd\xi}{2\pi}\,
\ln\bigl[1-e^{-2qd}\,r_1 r_2\bigr],
\label{eq:Eint}
\end{align}
where
\begin{equation}
r_\ell(q,i\xi)=\frac{v(q)\,\chi_{0,\ell}(q,i\xi)}
{1-v(q)\,\chi_{0,\ell}(q,i\xi)}\;\in(-1,0).
\label{eq:rdef}
\end{equation}
Here $\chi_{0,\ell}$ is the bare bubble, i.e., the proper (irreducible)
polarizability of layer $\ell$ at the RPA level, and the denominator
$1-v\chi_{0,\ell}$ generates the intralayer screening. Beyond RPA, the
quantity entering Eq.~\eqref{eq:rdef} is the proper polarizability
$\tilde\Pi_\ell$ of the isolated layer, which includes the intralayer
local-field (vertex) corrections; equivalently,
$r_\ell=v\chi_\ell$ with $\chi_\ell=\tilde\Pi_\ell/(1-v\tilde\Pi_\ell)$ the
full density response of the isolated layer. Inserting the full response
into Eq.~\eqref{eq:rdef} would count the intralayer screening twice.
Equation~\eqref{eq:Eint}, together with the capacitance formula
\eqref{eq:capdef}, is the principal result of this section: it is negative
(interlayer attraction), vanishes as $d\to\infty$, and its density Hessian
yields $\delta C^{-1}_{\rm inter}$. Replacing the bare bubbles
$\chi_{0,\ell}$ by the proper polarizabilities $\tilde\Pi_\ell$ of
the isolated layers extends Eq.~\eqref{eq:Eint} to the additive
approximation discussed in the next section; corrections to it are diagrams
in which interlayer interaction lines attach to a common vertex, interlayer
ladders, whose physical content is excitonic and whose role is examined in
Secs.~\ref{sec:neutrality} and \ref{sec:discussion}.

\section{Equivalence to the Lifshitz theory and the Casimir compressibility}
\label{sec:casimir}

\subsection{The identification argument}
\label{sec:identification}
The zero-temperature, nonretarded Lifshitz energy of two bodies facing each
other across a gap $d$, with specular reflection amplitudes
$r_{1,2}(q,i\xi)$ for evanescent waves of in-plane momentum $q$, is
precisely of the form \eqref{eq:Eint}~\cite{Lifshitz1956,Bordag2009,
Woods2016}. The identification is completed by showing that
Eq.~\eqref{eq:rdef} is the electrostatic reflection amplitude of a strictly
two-dimensional conducting sheet. Place the sheet at $z=0$ and apply an
external evanescent potential $\phi_{\rm ext}=\phi_0\,e^{iqx}e^{qz}$
incident from $z<0$. The sheet develops the induced density
$\delta n=\chi\,\phi_{\rm tot}(0)$, which radiates the potential
$\phi_{\rm ind}=v(q)\,\delta n\,e^{-q|z|}$. Solving the self-consistency
condition at $z=0$ gives $\phi_{\rm tot}(0)=\phi_0/(1-v\chi)$ and hence the
reflected amplitude
\begin{equation}
r(q,i\xi)=\frac{\phi_{\rm ind}}{\phi_{0}}\Big|_{z=0}
=\frac{v(q)\chi(q,i\xi)}{1-v(q)\chi(q,i\xi)},
\label{eq:refl}
\end{equation}
which is Eq.~\eqref{eq:rdef}. The response function $\chi$ in this
argument is defined with respect to the total potential at the sheet and is
therefore the proper polarizability; in terms of the full response
$\chi^{\rm full}=\chi/(1-v\chi)$ to the external potential, Eq.~\eqref{eq:refl}
reads simply $r=v\chi^{\rm full}$. A perfect metal corresponds to $r=-1$,
i.e., $\chi^{\rm full}\to-1/v$; a
weakly polarizable sheet to $r\simeq v\chi$. Thus the interlayer part of the bilayer ground-state energy is
the van der Waals-Casimir energy of the two sheets, and the interlayer
contribution to the inverse capacitance, Eq.~\eqref{eq:dcint}, is its second density derivative---a Casimir compressibility.

Capacitance measurements on double layers therefore probe a
dispersion force thermodynamically: Casimir experiments measure the force
$-\partial_d E_{\rm int}$~\cite{GarciaSanchez2012,Woods2016}, while a
capacitance bridge measures the second derivative
$\partial^2_{\delta n}E_{\rm int}$ of the same energy with respect to the
transferred charge.

\subsection{Domain of validity of the correspondence}
\label{sec:validity}
Three qualifications define where Eq.~\eqref{eq:Eint} applies.

(i)~\textit{Retardation.} The full Lifshitz theory replaces the electrostatic
kernel $e^{-2qd}$ by photon propagators; Eq.~\eqref{eq:Eint} is its
$c\to\infty$ limit. The nonretarded form holds while the relevant
fluctuation frequencies, $\xi\sim\omega_p(q\sim1/d)$ with
$\omega_p(q)=(2\pi ne^2q/m\bar\epsilon)^{1/2}$ the two-dimensional plasmon,
remain small compared with the photon frequency $cq$ at the dominant
momenta, $\omega_p(1/d)\ll c/d$. This condition fails only at
separations $d\gtrsim c^2m\bar\epsilon/2\pi ne^2$, which is macroscopic
(fractions of a millimeter and beyond) for all realistic densities; the
crossover to the true Casimir regime for two-dimensional sheets was mapped
in Ref.~\cite{SerneliusBjork1998}. Capacitors are therefore always
in the nonretarded regime, which is why plasmons, rather than photons,
play the role of the fluctuating modes here.

(ii)~\textit{Finite temperature.} At $T>0$ the frequency integral becomes a
Matsubara sum and Eq.~\eqref{eq:trlog} already has the required form. The
$\Omega_m=0$ term is the classical (static) fluctuation attraction; its
growing relative weight at high temperature is the bilayer analog of the
thermal Casimir force. All results below are at $T=0$.

(iii)~\textit{Additivity.} Lifshitz theory takes the response of each body as
given. Diagrammatically, Eq.~\eqref{eq:Eint} with the exact proper
polarizabilities $\tilde\Pi_\ell$ of the isolated layers resums all diagrams in which the interlayer lines connect closed
polarization loops; it omits interlayer vertex corrections
(particle-particle ladders), whose physical content is interlayer exciton
formation. These are negligible at high density and weak interlayer
coupling, but dominate in dilute bilayers and in electron-hole double
layers near exciton condensation, the same non-additivity that limits
Lifshitz theory for atomically close or chemically interacting
surfaces~\cite{Woods2016}.

\subsection{Relation to Coulomb drag}
\label{sec:drag}
The building blocks of Eq.~\eqref{eq:Eint}, the screened interlayer
interaction and the two polarization functions, are exactly those of the
Coulomb drag transresistivity~\cite{Narozhny2016,HwangDrag2011}. Drag
is determined by the dissipative part of the interlayer fluctuation
spectrum: it involves $\mathrm{Im}\,\chi_1\,\mathrm{Im}\,\chi_2$ at real
frequencies, weighted by a thermal factor. The Casimir compressibility is
determined by the reactive part of the same spectrum, an integral of the
coupled response over imaginary frequencies. The two observables thus probe
the same interlayer coupling, one out of equilibrium and the other in
equilibrium. This suggests quantitative
cross-checks between drag and capacitance data on the same device, a point
developed in Sec.~\ref{sec:discussion}.

\begin{figure*}[t]
\includegraphics[width=0.95\textwidth]{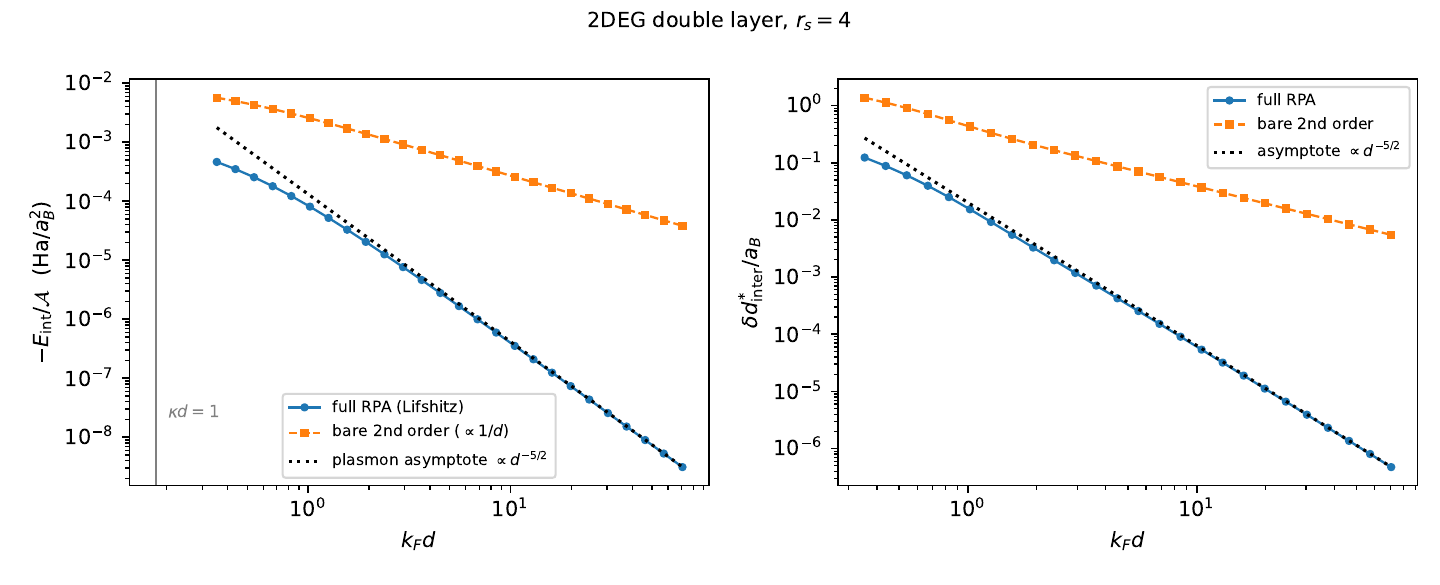}
\caption{2DEG double layer at $r_s=4$, equal densities. Left: interlayer
correlation energy from the full ring expression \eqref{eq:Eint} with the
exact Lindhard function, compared with bare second-order perturbation
theory ($\propto1/d$) and the plasmon zero-point asymptote
\eqref{eq:Easym} ($\propto d^{-5/2}$). Screening beyond $\kappa d\sim1$
suppresses the interaction by orders of magnitude. Right: the corresponding
positive shift of the effective capacitor thickness, Eq.~\eqref{eq:dstar},
converging onto the closed-form asymptote \eqref{eq:Kresult}, plotted
in the form of Eq.~\eqref{eq:dstarasym} (dotted line).}
\label{fig:2deg}
\end{figure*}

\section{Two-dimensional electron gas double layer}
\label{sec:2deg}

Throughout this section the layers are identical unpolarized 2DEGs and we
use effective atomic units. The Lindhard function on the imaginary axis has
the closed form [Appendix~\ref{app:lindhard}]
\begin{equation}
\chi_0(q,i\xi)=-\nu_0\Bigl[1-\tfrac1b\,\mathrm{Re}\sqrt{(b+ia)^2-1}\Bigr],
\label{eq:lindhard}
\end{equation}
with the dimensionless variables $b=q/2\kF$ and $a=\xi/qv_F$.

\subsection{High-density limit: a Gell-Mann-Brueckner analysis}
\label{sec:gmb}
Expanding Eq.~\eqref{eq:Eint} to second order in $v_{12}$ with bare bubbles
gives the leading interlayer term of the high-density expansion,
\begin{equation}
\frac{E_2}{\A}=-\frac{\hbar}{2}\int\!\frac{\dd^2q}{(2\pi)^2}\,v_{12}^2(q)
\int_{-\infty}^{+\infty}\frac{\dd\xi}{2\pi}\,\chi_{0,1}\chi_{0,2}.
\label{eq:E2}
\end{equation}
In the bulk electron gas the analogous second-order term diverges in the
infrared and forces the Gell-Mann--Brueckner ring
resummation~\cite{GellMannBrueckner1957}; in two dimensions the single-layer
second order is already finite~\cite{RajagopalKimball1977}, and the
interlayer term \eqref{eq:E2} is finite as well: the frequency integral
supplies a factor $\propto q$ at small $q$,
$\int_0^\infty\dd\xi\,[1-\xi/\sqrt{\xi^2+v_F^2q^2}]^2=(2-\tfrac\pi2)v_Fq$,
so the momentum integrand is regular. The high-density expansion
therefore terminates at second order; the ring resummation is needed only
when screening is strong (Sec.~\ref{sec:rings}). Moreover, because interlayer exchange is
absent (Sec.~\ref{sec:functional}), Eq.~\eqref{eq:E2} is the complete
leading term.

Scaling out the density, Eq.~\eqref{eq:E2} defines universal functions of
the single variable $x=\kF d$:
\begin{equation}
\frac{E_2}{N}=-\frac{2}{\pi}G(x)~\text{Ha},
\qquad
\delta C^{-1}_{2}=\frac{H(x)}{e^2 n},
\label{eq:GH}
\end{equation}
where $N=n\A$ is the electron number of one layer [per unit area,
$E_2/\A=-(2/\pi)\,nG(x)$~Ha]. Normalized per particle, the energy carries
no residual $r_s$ dependence in absolute units, the bilayer
counterpart of the Gell-Mann-Brueckner constant. Explicitly, with
$f(b,a)=1-\mathrm{Re}\sqrt{(b+ia)^2-1}/b$ the dimensionless Lindhard
function of Eq.~\eqref{eq:lindhard}, $\chi_0=-f/\pi$, and
$\partial_n=(\pi/\kF)\partial_{\kF}$ acting on $\chi_0$ through $b$ and $a$,
\begin{align}
G(x)&=2\int_0^\infty\!\dd b\,e^{-4xb}\!\int_0^\infty\!\dd a\,f^2(b,a),
\nonumber\\
H(x)&=-n\!\int\!\frac{\dd^2q}{(2\pi)^2}\,v_{12}^2(q)\!
\int_{-\infty}^{\infty}\!\frac{\dd\xi}{2\pi}\,
\bigl[\chi_0\,\partial_n^2\chi_0-(\partial_n\chi_0)^2\bigr],
\label{eq:GHdef}
\end{align}
the latter evaluated at $n_1=n_2=n$ in effective atomic units; both depend
on $\kF d$ only. Numerically $G(0)=0.964$, and since
$f(0,a)=1-a/\sqrt{1+a^2}$ the tails are
$G(x)\to(2-\pi/2)/2x$ [equivalently $E_2/\A\to-(2-\tfrac\pi2)
e^4\nu_0^2v_F/2d$] and $H(x)\simeq0.097/x$, and $H(x)>0$ everywhere:
already at second order, interlayer correlations reduce the capacitance at
high density. In terms of the effective thickness,
$\delta d^*_2=\tfrac{r_s^2}{4}H(\kF d)\,\aB$.

\subsection{Ring resummation and plasmon zero-point energy}
\label{sec:rings}
At fixed $\kF d$ the parameter controlling successive rings is
$\kappa d=\sqrt2\,r_s(\kF d)$. For $\kappa d\gtrsim1$ the full logarithm of
Eq.~\eqref{eq:Eint} must be kept, and screening converts the bare $1/d$ tail
of $E_2$ into a much faster decay. For $\kappa d\gg1$ the integral is
dominated by $q\sim1/2d\ll\{\kappa,\kF\}$, where
$r(q,i\xi)\simeq-\omega_p^2(q)/[\omega_p^2(q)+\xi^2]$ with
$\omega_p^2(q)=2\pi ne^2q/m\bar\epsilon$. The frequency integral is then
elementary and Eq.~\eqref{eq:Eint} becomes the change in zero-point energy
of the coupled optical and acoustic plasmons of the bilayer,
\begin{equation}
\frac{E_{\rm int}}{\A}
=\frac{\hbar}{4\pi}\int_0^\infty\!\dd q\,q
\bigl[\omega_+(q)+\omega_-(q)-2\omega_p(q)\bigr],
\label{eq:zeropoint}
\end{equation}
with $\omega_\pm=\omega_p\sqrt{1\pm e^{-qd}}$. Equation~\eqref{eq:zeropoint}
makes the analogy with the Casimir effect explicit: the photon modes of the
metallic Casimir problem are replaced by the two-dimensional plasmons.
Substituting $u=qd$,
\begin{equation}
\frac{E_{\rm int}}{\A}
=\frac{\hbar}{4\pi}\sqrt{\frac{2\pi ne^2}{m\bar\epsilon}}\,\frac{J}{d^{5/2}},
\quad
J=-0.0629780,
\label{eq:Easym}
\end{equation}
where $J=\int_0^\infty\dd u\,u^{3/2}[\sqrt{1+e^{-u}}+\sqrt{1-e^{-u}}-2]$
admits the rapidly convergent series
$J=\tfrac{3\sqrt\pi}{2}\sum_{k\ge1}\binom{1/2}{2k}(2k)^{-5/2}$
[Appendix~\ref{app:series}].

This $d^{-5/2}$ law is the fractional-power van der Waals attraction
of two-dimensional metals known from the theory of dispersion forces.
Equation~\eqref{eq:Easym} reproduces, including the coefficient, the
numerical result of Sernelius and Bj\"ork for a pair of quantum wells,
$E\simeq-0.012\,562\,e\hbar\sqrt{n/\bar\epsilon m^*}\,d^{-5/2}$
\cite{SerneliusBjork1998}, since $|J|\sqrt{2\pi}/4\pi=0.012\,563$. The
fractional powers of thin metallic films were analyzed in
Refs.~\cite{Barash1989,BostromSernelius2000}, and Dobson, White, and Rubio
established $d^{-5/2}$ (two-dimensional metals) and $d^{-3}$ (undoped
graphene) as analytic benchmarks that pairwise-additive van der Waals
functionals fail to reproduce~\cite{Dobson2006}; the graphene case at RPA
level was studied in Ref.~\cite{Sarabadani2011}. Figure~\ref{fig:2deg}
(left) shows the full evaluation of Eq.~\eqref{eq:Eint} with the exact
Lindhard function crossing over from the second-order $1/d$ behavior at
$\kappa d\lesssim1$ to the plasmon law \eqref{eq:Easym}, which it approaches
to within $2\%$ by $\kF d\simeq30$.

\subsection{Capacitance correction: closed asymptote, sign, decomposition}
\label{sec:capresult}
Differentiating Eq.~\eqref{eq:zeropoint} at fixed total density, with
$\omega^2_{p,\ell}\propto q\,n_\ell$, gives the central quantitative result
for parabolic layers [Appendix~\ref{app:series}]:
\begin{equation}
\delta C^{-1}_{\rm inter}
=\frac{\hbar}{4\pi e^2n^2}
\sqrt{\frac{2\pi ne^2}{m\bar\epsilon}}\,
\frac{K}{d^{5/2}},
\qquad K=+0.0477118,
\label{eq:Kresult}
\end{equation}
where $K=\int_0^\infty\dd u\,u^{3/2}D(e^{-u})$ with
\begin{equation}
D(a)=\frac12-\frac{\sqrt{1-a^2}}{\sqrt{1+a}+\sqrt{1-a}}
=\frac{3}{16}a^2+\frac{17}{256}a^4+\dots
\label{eq:Dfun}
\end{equation}
The compact form of $D$ makes the sign manifest: since
$\sqrt{1-a^2}=\sqrt{(1+a)(1-a)}$, the arithmetic--geometric-mean inequality
gives $D(a)\ge0$ on $0\le a\le1$. Throughout the high-density regime,
interlayer ring correlations increase $1/C$: the van der Waals binding of
the layers is maximal at equal densities, so transferring charge costs
correlation energy and opposes charging. (The curvature integrand and
its numerically stable form are derived in Appendix~\ref{app:series};
quadrature details are given in Appendix~\ref{app:numerics}.) In
dimensionless variables,
\begin{equation}
\delta d^*_{\rm inter}\simeq 0.01\,
\frac{\sqrt{r_s}}{(\kF d)^{5/2}}\;\aB,
\qquad \kappa d\gg1.
\label{eq:dstarasym}
\end{equation}
Equations~\eqref{eq:Easym} and \eqref{eq:Kresult} are two derivatives
of the same functional, and in the plasmon regime they are related directly.
Both scale as $d^{-5/2}$, so the attractive Casimir pressure between the
layers, $P_C\equiv|\partial_d E_{\rm int}|/\A=\tfrac52|E_{\rm int}|/\A d$,
determines the capacitance correction with a universal coefficient,
\begin{equation}
\delta C^{-1}_{\rm inter}=\frac{K}{|J|}\,\frac{|E_{\rm int}|}{\A\,e^2n^2}
=\frac{2K}{5|J|}\,\frac{d\,P_C}{e^2n^2},
\qquad \frac{2K}{5|J|}=0.303.
\label{eq:forcecap}
\end{equation}
The same relation holds for Dirac layers with $K\to K_{1/2}$
[$2K_{1/2}/5|J|=0.126$; Sec.~\ref{sec:mlg}]. Equation~\eqref{eq:forcecap}
shows in what sense a capacitance measures the same dispersion force
as a Casimir experiment: when the plasmon zero-point motion dominates, the
Casimir pressure determines the interlayer capacitance correction, and vice
versa, without free parameters. Away from
this regime the two observables remain derivatives of the single functional
of Eq.~\eqref{eq:Eint}, but no closed relation between them exists.
The full ring evaluation converges onto Eq.~\eqref{eq:Kresult} [ratio
$0.94$ at $\kF d=28$, $0.99$ at $\kF d=113$; Fig.~\ref{fig:2deg}, right].
The decomposition of the curvature is instructive: the mixed-derivative
(``cross'') part $-2\partial_{n_1}\partial_{n_2}\eps_{\rm int}$ contributes
the fraction $0.335$ of Eq.~\eqref{eq:Kresult}, essentially independent of
$d$ and $r_s$ (the plasmon-pole model gives $0.3345$ analytically, and
exactly $\tfrac13$ at leading order in $e^{-qd}$). The remaining two thirds
is the $d$-dependent renormalization of each layer's compressibility by the
presence of the other layer. 

To estimate the magnitude: in GaAs double quantum wells
($\aB\simeq10$~nm, $r_s\simeq2$, $d\simeq30$~nm, $\kF d\simeq2$) the
second-order estimate gives $\delta d^*\sim\tfrac{r_s^2}{4}H(2)\simeq
0.05\,\aB\simeq0.5$~nm, reduced further by screening toward
Eq.~\eqref{eq:dstarasym}, a percent-level, density- and
separation-dependent correction to $d^*$, distinguishable from the
geometric term ($\propto d$, density independent) and from the single-layer
compressibility terms ($d$ independent) by its $n^{-3/2}d^{-5/2}$
signature.

The second-order
functions $G$ and $H$ are controlled for $r_s\ll1$. The large-separation
asymptotes, Eqs.~\eqref{eq:Easym}, \eqref{eq:Kresult}, and
\eqref{eq:forcecap}, are exact beyond the ring approximation at leading
order in $1/\kappa d$, for any $r_s$ at which the layer is a Fermi liquid:
for $\kappa d\gg1$ the integral is dominated by $q\ll\kF$ and
$\xi\sim\omega_p\gg v_Fq$, where the proper polarizability of a parabolic
layer is fixed by the $f$-sum rule, $\tilde\Pi\to-nq^2/m\xi^2$, so that the
reflection amplitude reduces to the exact long-wavelength plasmon pole and
the constants $J$ and $K$ do not depend on the approximation used for the
layer response; $r_s$ enters only through the prefactor $\sqrt n$. The ring
approximation is used at RPA accuracy only in the crossover region
$\kappa d\sim1$ and in the rate of approach to the asymptote. There the
omitted intralayer local-field corrections to $\tilde\Pi$ are of relative
order $G_{\rm lf}(q\sim1/d)\sim1/\kF d$, where $G_{\rm lf}(q)$ is the
static local-field factor of the layer, which vanishes linearly as
$q\to0$. The value $r_s=4$ in
Fig.~\ref{fig:2deg} was chosen so that the crossover at $\kappa d=1$ and the
asymptotic regime both fall in the plotted window; the curves illustrate the
crossover at the RPA level.

\begin{figure*}[t]
\includegraphics[width=0.95\textwidth]{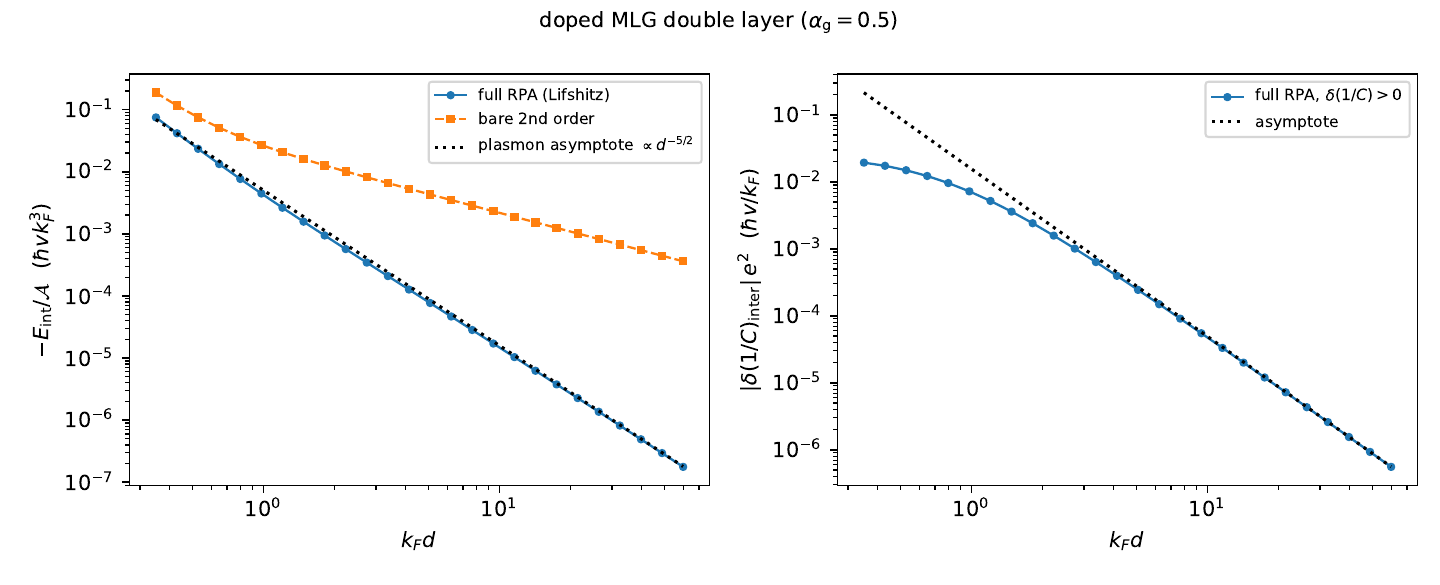}
\caption{Two parallel sheets of doped monolayer graphene, $\alpha_{\rm g}=0.5$, equal
densities, natural units $\hbar=v=1$. Left: interlayer correlation energy
(full ring expression, bare second order, plasmon zero-point asymptote).
Right: interlayer capacitance correction, $e^2\delta C^{-1}_{\rm
inter}$ in units of $\hbar v/\kF$, positive throughout, converging
onto the closed asymptote \eqref{eq:mlgasym}. The reference density
$n=10^{12}$~cm$^{-2}$ corresponds to $\kF^{-1}=5.7$~nm.}
\label{fig:mlg}
\end{figure*}

\section{Graphene double layers}
\label{sec:graphene}

We now evaluate the theory for doped MLG and BLG double layers using the
full dynamical polarizabilities. Band models: MLG,
$\eps_\pm(k)=\pm\hbar vk$, chirality $J=1$; BLG, the two-band model
\cite{McCann2006}, $\eps_\pm(k)=\pm\hbar^2k^2/2m$, $J=2$; coherence factors
$F_{ss'}(\bm k,\bm k')=\tfrac12[1+ss'\cos J\varphi_{\bm k\bm k'}]$, $g=4$.
Reference parameters: $\alpha_{\rm g}=0.5$ (hBN encapsulation) for MLG;
$\kF\aB^*=1.25$ for BLG ($\aB^*=\bar\epsilon\hbar^2/me^2\simeq7$~nm for
$m=0.033m_0$, $\bar\epsilon=4.4$), corresponding to
$n=10^{12}$~cm$^{-2}$, $\kF=0.177$~nm$^{-1}$.

\subsection{Polarizabilities for MLG and BLG devices}
\label{sec:polarizability}
The dynamical polarizabilities of doped MLG and BLG are available in closed
form~\cite{Wunsch2006,HwangDasSarma2007,Ramezanali2009,Sensarma2010,
Gamayun2011} and underlie the Coulomb-drag theory of these
systems~\cite{HwangDrag2011}. The published expressions are lengthy and
piecewise and do not allow analytical progress.
We therefore compute the $T=0$ imaginary-axis bubbles directly from
the band models~\footnote{In this subsection we set $\hbar=1$, so that the
imaginary frequency $\xi$ is measured in energy units.},
\begin{align}
\Pi(q,i\xi)&=\Pi_{\rm vac}(q,i\xi)
\nonumber\\
&+g\sum_{s'=\pm}\int_{k<\kF}\!\!\frac{\dd^2k}{(2\pi)^2}
F_{+s'}(\bm k,\bm k+\bm q)\,
\frac{2\Delta_{s'}}{\xi^2+\Delta_{s'}^2},
\label{eq:bubble}
\end{align}
$\Delta_{s'}=\eps_+(k)-\eps_{s'}(|\bm k+\bm q|)$, with the intrinsic parts
$\Pi^{\rm MLG}_{\rm vac}=-\tfrac{g}{16}q^2/\sqrt{v^2q^2+\xi^2}$ (exact)
and $\Pi^{\rm BLG}_{\rm vac}=-N_0\,\psi(2m\xi/q^2)$ computed
numerically [$N_0=gm/2\pi$]. The dimensionless function $\psi$ is the
interband bubble of the filled valence band in units of $N_0$, written in
the scaled variables $\bm k=q\bm\kappa$ and $z=2m\xi/q^2$:
\begin{equation}
\psi(z)=\frac1\pi\int\!\dd^2\kappa\,
\frac{\bigl[1-\cos2\varphi_{\bm\kappa}\bigr]\,\delta(\bm\kappa)}
{z^2+\delta^2(\bm\kappa)},
\quad
\delta(\bm\kappa)=\kappa^2+|\bm\kappa+\hat{\bm e}|^2,
\label{eq:psi}
\end{equation}
where $\hat{\bm e}$ is the unit vector along $\bm q$ and
$\varphi_{\bm\kappa}$ is the angle between $\bm\kappa$ and
$\bm\kappa+\hat{\bm e}$. The integral converges because the $J=2$
coherence factor $1-\cos2\varphi_{\bm\kappa}$ vanishes as $\kappa^{-2}$ at
large $\kappa$; $\psi(0)=\ln4$ reproduces the static intrinsic
result~\cite{HwangKohn2008}, and $\psi$ decreases monotonically. The evaluation is checked
against all limits on which the published results agree: 
the MLG static response \cite{HwangDasSarma2007}, the
hydrodynamic limit, the intrinsic-BLG constant $N_0\ln4$
\cite{HwangKohn2008}, the compressibility sum rule
$\Pi(q\to0,0)=-N_0$ of doped BLG together with the known shape of its
static response (flat, then a peak $1.75N_0$ at $q=2\kF$, decaying to
$N_0\ln4$)~\cite{HwangKohn2008,Sensarma2010}, and $f$-sum-rule tails. 
At this level, our results are consistent with those reported in Refs.~\cite{Wunsch2006,HwangDasSarma2007,Ramezanali2009,
Sensarma2010,Gamayun2011,HwangDrag2011,Profumo2010,Sarabadani2011}.

One technical point should be noted. The static limit of
Eq.~\eqref{eq:bubble} contains principal-value structures at the
Fermi-surface crossing angle and, for BLG only, at the backscattering
configuration $k=q/2$, $\theta=\pi$, where the $J=2$ coherence factor is
maximal, whereas the $J=1$ factor of MLG vanishes. This is the same
matrix-element structure responsible for the absence of backscattering
suppression in BLG transport; numerically it makes the BLG bubble
considerably harder to evaluate than the MLG one (Appendix~\ref{app:numerics}).

For doped graphene, the justification for using the RPA does not come from a 
small coupling constant: $\alpha_{\rm g}$ is independent of density and can be of 
order unity. Rather, the approximation relies on the fourfold flavor degeneracy 
of graphene ($N=g=4$), which makes the ring diagrams the leading contribution 
in a $1/N$ expansion~\cite{Son2007,FosterAleiner2008,Kotov2012}. 
This provides the rationale for using the RPA even at 
$\alpha_{\rm g}\sim0.5$--$2$, as we do throughout this section.

\subsection{Doped monolayer graphene}
\label{sec:mlg}
Figure~\ref{fig:mlg} shows the interlayer energy and capacitance correction
for two parallel sheets of doped monolayer graphene. The
behavior is qualitatively the same as for the electron gas. The energy crosses over from the bare second-order
behavior to the plasmon zero-point law, now with
$\omega_p^2(q)=ge^2\mu q/2\bar\epsilon\hbar^2$, approached from below with
a characteristic $\sim15\%$ interband suppression around $\kF d\sim1$--$2$.
The capacitance correction is positive at all separations studied
($0.35\le\kF d\le60$), with the closed asymptote
\begin{equation}
\delta C^{-1}_{\rm inter}
=\frac{\hbar}{4\pi e^2n^2}
\sqrt{\frac{ge^2\mu}{2\bar\epsilon\hbar^2}}\,
\frac{K_{1/2}}{d^{5/2}},
\qquad K_{1/2}=0.0198002,
\label{eq:mlgasym}
\end{equation}
the $\gamma=\tfrac12$ member of a one-parameter family of constants
$K_\gamma$ obtained by repeating the zero-point analysis with
$\omega^2_{p,\ell}\propto q\,n_\ell^\gamma$ [Appendix~\ref{app:series}];
$\gamma=\tfrac12$ because $\omega_p^2\propto\mu\propto\sqrt n$ for Dirac
carriers. The computed curvature reaches $0.99$ of Eq.~\eqref{eq:mlgasym}
by $\kF d=30$. The coefficient is smaller than the parabolic value
$K=K_1$ by the factor $2.4$: the Dirac plasmon is less sensitive to density
transfer. At $n=10^{12}$~cm$^{-2}$ and $d=5.7$~nm the correction amounts
to $\delta d^*_{\rm inter}\simeq6.5\times10^{-3}$~nm, about $0.2\%$ of the
kinetic quantum-capacitance thickness of the two layers, with the
characteristic scaling $\delta C^{-1}\propto n^{-7/4}d^{-5/2}$.
Beyond the RPA, the long-wavelength plasmon weight of a Dirac liquid is
not protected by Galilean invariance: the Drude weight $D$ is renormalized
by interactions, $D/D_0\simeq1.1$--$1.2$ at $\alpha_{\rm g}\sim0.5$ and
typical densities~\cite{Abedinpour2011}.
Since Eq.~\eqref{eq:mlgasym} depends on the plasmon weight only through
$\sqrt{\mu}$, this correction is absorbed by $\mu\to\mu D/D_0$ and shifts
the asymptotic coefficient by $5$--$10\%$; the constant $K_{1/2}$ itself
is unaffected.

\begin{figure*}[t]
\includegraphics[width=0.95\textwidth]{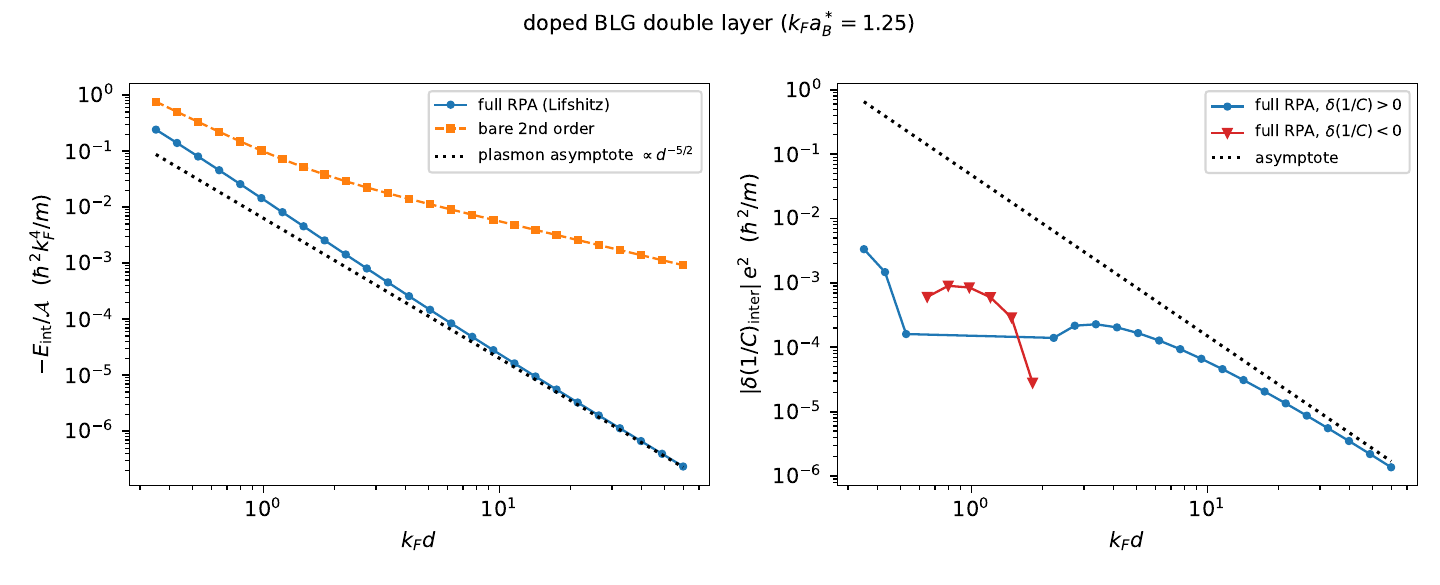}
\caption{Two parallel sheets of doped Bernal bilayer graphene (two-band
model),
$\kF\aB^*=1.25$, equal densities, natural units $\hbar=m=1$. Left:
interlayer correlation energy. Right: magnitude of the interlayer
capacitance correction, $|e^2\delta C^{-1}_{\rm inter}|$ in units of
$\hbar^2/m$; downward triangles mark the window
$0.55<\kF d<1.87$ where the correction is negative (capacitance enhanced).
The dotted line is the parabolic-band asymptote, approached slowly because
of interband screening.}
\label{fig:blg}
\end{figure*}

\subsection{Doped bilayer graphene: interband suppression and sign reversal}
\label{sec:blg}
Two parallel sheets of doped bilayer graphene behave differently
from the electron gas despite the
parabolic bands (Fig.~\ref{fig:blg}). 
Three observations can be drawn from the numerical analysis described in Appendix~\ref{app:numerics}:

(i)~\textit{Strong suppression.} A control calculation for a fictitious
four-flavor parabolic gas with the same mass, density, and interaction
(BLG stripped of interband transitions and chirality factors) yields a
positive curvature $30$--$100$ times larger than the true two-band result
over $0.5\le\kF d\le4$. The BLG interlayer correction is the small residue
of a near-cancellation driven by interband screening: even undoped BLG
responds with the finite constant $N_0\ln4$~\cite{HwangKohn2008}, so each
sheet acts as an almost frequency-flat mirror, $r\to-1$, over a wide
$(q,\xi)$ domain, and a capacitor formed by two nearly perfect mirrors has
a Casimir energy nearly independent of the charge transfer.

(ii)~\textit{Sign-reversal window.} The residue changes sign twice: at
$\kF\aB^*=1.25$,
\begin{equation}
\delta(1/C)_{\rm inter}<0
\quad\text{for}\quad
0.55\lesssim\kF d\lesssim1.87,
\label{eq:window}
\end{equation}
i.e., for $3~\text{nm}\lesssim d\lesssim11$~nm at $n=10^{12}$~cm$^{-2}$,
which is the range typical of experiments, interlayer correlations weakly
enhance the capacitance of a BLG double layer, opposite in sign to the
2DEG and MLG results and obtained here already at ring level, without
invoking excitonic physics. Outside the window the correction is positive
and approaches the parabolic asymptote [Eq.~\eqref{eq:Kresult} with the BLG
mass] slowly ($80\%$ at $\kF d=60$).

(iii)~\textit{Smallness of the correction.} In the window
$|\delta d^*_{\rm inter}|\sim5\times10^{-4}$~nm at $10^{12}$~cm$^{-2}$:
within the ring approximation, interlayer correlations are for practical
purposes absent from the capacitance of doped BLG double layers.

One should keep in mind, however, that the residual identified in 
(i) is only about $1\%$ of the individual contributions that nearly cancel. 
Its value may therefore be affected by corrections beyond the two-band model, 
including the four-band structure at $\eps_F\gtrsim\gamma_1/4$, 
trigonal warping at very low energies, and, in particular, the gap induced 
by the displacement field generated by the charging itself~\cite{McCann2006,Young2012}.
The two statements above have different status. The suppression
requires only that the interband polarizability of BLG be comparable to the
intraband one ($N_0\ln4$ versus $N_0$), which makes each sheet an almost
frequency-independent mirror; this is a low-energy consequence of the
chirality-two band contact and persists in the four-band model for
$\eps_F\ll\gamma_1/4$, so the suppression is robust. The sign of the
residue, by contrast, is decided by the balance of two nearly equal
contributions and can be changed by perturbations of the spectrum at the
percent level. The precise boundaries, and even the existence, of the
sign window should therefore be regarded as two-band-model statements
pending a four-band, self-consistently gapped calculation.

\subsection{Limits of validity near neutrality}
\label{sec:neutrality}
The central formula, Eq. \eqref{eq:Eint}, relies on assumptions that become 
invalid as either layer approaches its neutrality point, a regime that capacitance experiments routinely probe.

(i)~\textit{Loss of the controlled expansion.} For the 2DEG the ring sum
is exact as $r_s\to0$. For MLG the coupling $\alpha_{\rm g}$ is density
independent, and screening weakens as $q_{TF}=g\alpha_{\rm g}\kF\to0$; for
BLG the effective $r_s\propto1/\kF$ diverges at neutrality. Below
$|n|\sim10^{11}$~cm$^{-2}$ the ring approximation is uncontrolled.

(ii)~\textit{Relative weight of the interlayer term.} At fixed
separation $d$ and decreasing density, the ratio of the interlayer
correction to the kinetic quantum capacitance stays below the percent level
for MLG down to $\kF d\simeq0.1$ (both quantities grow together as
$n\to0$, $\propto n^{-1/2}$), whereas for BLG, whose kinetic term is density
independent, the ratio grows monotonically as $n\to0$. No small
parameter justifies a description of BLG double layers near neutrality in
terms of single-layer quantities alone.

(iii)~\textit{The excitonic channel.} The capacitance curvature displaces
the system to $(\bar n+\delta n,\bar n-\delta n)$; at double neutrality
$\bar n=0$ this is an electron-hole double layer, whose interlayer
particle-particle channel is the pairing channel of exciton
condensation~\cite{LozovikYudson1976,Min2008,Perali2013}, observed in
quantum-Hall double bilayers and in strongly enhanced interlayer tunneling
at total neutrality~\cite{LiDean2017,Burg2018}. Near the condensation the
imbalance curvature acquires a negative contribution that rings miss
entirely (in the language of Ref.~\cite{Berthod2021}, the interlayer
polarizability diverges for bound interlayer pairs): the capacitance at
double neutrality is an order-parameter susceptibility, not a
density-of-states measurement. Equation~\eqref{eq:Eint} fails there for
the same reason Lifshitz theory fails for chemically bonding surfaces.

(iv)~\textit{Disorder.} Electron--hole puddles smear
$|n|\lesssim10^{10}$--$10^{11}$~cm$^{-2}$~\cite{Martin2008}.

(v)~\textit{BLG band-structure limits.} Two-band validity requires
$\eps_F\ll\gamma_1/4$; trigonal warping invalidates the $J=2$ chirality at
meV scales; and the charging field gaps the spectrum.

\section{Discussion: limiting regimes, prior results, and the crossover}
\label{sec:discussion}

We now summarize the behavior of $\delta C^{-1}_{\rm inter}$ across the bilayer phase diagram, 
focusing on the two limiting regimes where controlled results are available and discuss the crossover between them.

(i) \textit{High density (this work).} The second-order functions
$G,H$ at $\kappa d\lesssim1$ are controlled for $r_s\ll1$, and the plasmon
zero-point asymptote \eqref{eq:Kresult} at $\kappa d\gg1$ is exact at
leading order in $1/\kappa d$ for any $r_s$ (Sec.~\ref{sec:capresult}).
The sign is positive: the correction
is carried by the particle-hole (plasmon) channel, and transferring charge
weakens the van der Waals binding, which is maximal at equal densities.

(ii) \textit{Low density.} The opposite limit is classical and well
understood. The line of thought goes back to Bello, Levin, Shklovskii, and
Efros~\cite{Bello1981} and to Chenskii and Tkach~\cite{ChenskiiTkach1984},
who showed that Coulomb correlations among localized surface charges
produce a negative effective density of states in capacitance. This was the
earliest example of correlation-dominated capacitance and the conceptual
ancestor of the negative compressibility measured in
Refs.~\cite{Kravchenko1990,Eisenstein1992,Eisenstein1994}. For a dilute
double layer, $r_s\gg1$ and $nd^2\lesssim1$, the electrons form
interpenetrating Wigner-crystal-like configurations; a transferred electron
remains bound to the correlation hole it leaves behind, and the resulting
electron-hole dipoles resist charging only through their weak mutual
repulsion. The capacitance is then anomalously large,
$d^*\to0$~\cite{SkinnerShklovskii2010}, consistent with the large
enhancement observed in Ref.~\cite{LiAshoori2011}; equivalently, the
interlayer contribution to $1/C$ is large and negative, cancelling most of
the geometric term. Even at equal densities the correlation hole is visible
in the energetics: when each electron faces a depleted disk in the opposite
layer, the interlayer energy grows linearly with $d$, $E_{\rm
int}(d)/\A\simeq E_{\rm int}(0)/\A+2\pi e^2n_1n_2d/\bar\epsilon$, the
electrostatic energy of a double layer of correlation holes. [For a fixed
interlayer pair correlation $g_{12}(r)$ and $d\ll n^{-1/2}$ the linear term
is exactly $(2\pi e^2/\bar\epsilon)\,n_1n_2[1-g_{12}(0)]\,d$, since
$\int\dd^2r\,[(r^2+d^2)^{-1/2}-r^{-1}]=-2\pi d$]. This is the
microscopic origin of the capacitance enhancement: with $n_1n_2=\bar n^2-\delta n^2$ its
curvature $\partial^2_{\delta n}$ equals $-4\pi e^2d/\bar\epsilon$ and
cancels the geometric term of Eq.~\eqref{eq:decomp} identically, so the
$d^*\to0$ of Ref.~\cite{SkinnerShklovskii2010} follows exactly in the
correlation-hole picture. The variational treatment
of gate-screened layers~\cite{SkinnerFogler2010} and the bilayer
Wigner-crystal phase diagram~\cite{Goldoni1996} refine this limit.

(iii) \textit{The crossover.} The sign of $\delta C^{-1}_{\rm inter}$ must
therefore change along a line in the $(r_s,\kF d)$ plane. In diagrammatic
language this is a competition between the particle-hole channel (rings;
positive contribution, dominant at high density) and the
particle-particle channel (interlayer excitons and dipoles; negative
contribution, dominant at low density). These are the two mechanisms
identified at lowest order in the cross-quantum-capacitance analysis of
Berthod \emph{et al.}~\cite{Berthod2021}, now supported by controlled
calculations in both limits. Their perturbative treatment of the
particle--particle channel produces a logarithmic singularity at equal
layer densities, the weak-coupling signature of a two-dimensional interlayer
bound state; the singularity is cut by the exponentially small exciton
binding energy, which is why this channel is weak, and the ring result
controlled, at high density. In our view, locating the sign-change
line is the most concrete problem that this physics poses for quantum Monte
Carlo: it requires $E(\bar n+\delta n,\bar n-\delta n;d)$ to order $\delta n^2$ at a
handful of $(r_s,\kF d)$ points, an imbalance-resolved bilayer
calculation that existing bilayer QMC data (taken at fixed equal densities
or fixed polarization~\cite{Marchi2009,GoriGiorgi2004}) do not provide.
The Singwi-Tosi-Land-Sj\"olander treatment of interlayer local
fields~\cite{Zheng1994,Swierkowski1993,Moudgil2002}, benchmarked at small
$r_s$ against the exact $H(x)$, is the natural interpolation tool in the
meantime; Hartree-Fock studies of double layers~\cite{Hanna2000} capture
the exchange-driven part of the crossover.

(iv) \textit{BLG anomaly.} Bilayer graphene falls outside this
classification: there
the ring contribution itself is suppressed by interband screening and
reverses sign in the accessible window, Eq.~\eqref{eq:window}. For
experiments of the type of Refs.~\cite{Henriksen2010,Young2012,Zibrov2017}
the implication is that interpreting BLG double-layer capacitance in terms
of single-layer compressibilities alone is quantitatively safe at doped
densities. This is not because interlayer correlations are weak, but
because the dispersion energy of two nearly perfect mirrors is insensitive
to their charge dynamics.

(v) \textit{Drag-capacitance correspondence.} Drag and the Casimir
compressibility are two moments of the same interlayer fluctuation spectrum
(Sec.~\ref{sec:drag}). Within the ring approximation both follow from the
same screened interlayer propagator, so a model that fits the measured
drag~\cite{Gramila1991,Gorbachev2012} predicts $\delta C^{-1}_{\rm inter}$
with no adjustable parameters. Near the exciton instability both quantities
become anomalous together, perfect drag~\cite{Narozhny2016,LiDean2017} and
the capacitance anomaly of Sec.~\ref{sec:neutrality}; formulating exact sum
rules linking the two is an open problem.

(vi) \textit{Experimental outlook.} The interlayer correction carries a
parameter-free signature: $\delta C^{-1}_{\rm inter}\propto
n^{-3/2}d^{-5/2}$ with coefficient \eqref{eq:Kresult} for parabolic
layers, and $n^{-7/4}d^{-5/2}$ with coefficient \eqref{eq:mlgasym} for
MLG. It is separable from the geometric term (density independent,
$\propto d$) and from the single-layer compressibilities ($d$ independent,
directly measurable in the penetration-field
geometry~\cite{Eisenstein1992,Eisenstein1994}) by varying $n$ at several
fixed $d$ in one heterostructure family. The predicted magnitudes are
small: at the percent level of $d^*$ in GaAs double wells, and $10^{-3}$ of
the single-layer quantum capacitance in graphene at $10^{12}$~cm$^{-2}$.
Two points bear on their observability. First, the geometric term is not a
background of unknown size: $d$ is measured independently, and capacitance
spectroscopy of correlations, from the negative compressibility of
Refs.~\cite{Eisenstein1992,Eisenstein1994} onward, has consisted of resolving
percent-level shifts of $d^*$ against it, while modern techniques resolve
considerably smaller relative signals~\cite{Zibrov2017,Zhou2021}. Second,
the interlayer term is the only contribution to $d^*$ that depends on $n$
and $d$ jointly, so a family of devices at several $d$, or the
penetration-field geometry at fixed $d$ and several $n$, isolates it without
a subtraction model. In this sense it is, like Coulomb drag, determined
exclusively by interlayer correlations: it vanishes identically without
them. The effect grows rapidly toward low density and small $\kF d$, where
the crossover physics discussed above takes over. A measurement of $\delta
d^*_{\rm inter}$ would constitute a thermodynamic detection of a dispersion
force.

\section{Summary}
\label{sec:summary}

We have developed a theory of the interlayer-correlation contribution to
the quantum capacitance of electron double layers. The main results are as
follows.

(1)~Within the functional-integral (RPA) framework, the
separation-dependent part of the bilayer ground-state energy is exactly the
nonretarded Lifshitz energy of two conducting sheets,
Eqs.~\eqref{eq:Eint}--\eqref{eq:refl}, with reflection amplitudes built
from the layer polarizabilities. The interlayer capacitance correction,
Eq.~\eqref{eq:dcint}, is the second density derivative of this dispersion
energy (a Casimir compressibility). Retardation is negligible for all
capacitor geometries; the zero-point modes are the coupled plasmons of the
bilayer.

(2)~For the 2DEG double layer the high-density limit is solved: the
interlayer second-order energy is infrared finite (no Gell-Mann-Brueckner
resummation is forced in two dimensions) and defines universal functions
$G(\kF d)$, $H(\kF d)$; at $\kappa d\gg1$ the ring sum
is dominated by plasmon zero-point motion and yields the closed asymptotes
\eqref{eq:Easym} and \eqref{eq:Kresult} with universal constants
$J=-0.0629780$ and $K=0.0477118$, the former reproducing the quantum-well
dispersion energy of Ref.~\cite{SerneliusBjork1998} to all quoted digits.
The sign is proven positive: at high density, interlayer correlations
oppose charging. The mixed-derivative (cross) term is only one third of
the effect.

(3)~For graphene: doped monolayers behave like the electron gas with
$K_{1/2}=0.0198002$; doped bilayer graphene is anomalous, interband
screening suppresses the interlayer correction by one to two orders of
magnitude and, within the two-band model, reverses its sign in the window
$0.55\lesssim\kF d\lesssim1.9$, Eq.~\eqref{eq:window}. The suppression
is robust; the sign window is a two-band-model result whose boundaries, and
existence, are sensitive to the displacement-field gap and to trigonal
warping (Sec.~\ref{sec:blg}).

(4)~The theory has well-defined limits of validity. Upon dilution the particle--particle
(interlayer-exciton) channel, absent from the ring diagrams, grows and acts
in the opposite sense, making the correction large and
negative~\cite{SkinnerShklovskii2010,LiAshoori2011}; a line in the
$(r_s,\kF d)$ plane must therefore separate capacitance reduction from
capacitance enhancement. No resummation is controlled in its vicinity, and
we pose the location of this sign-change line as a concrete target for
imbalance-resolved bilayer quantum Monte Carlo. The end point of this
evolution is reached in electron-hole double layers beyond the
exciton-condensation transition: the transferred charge then measures the
equation of state of the dipolar exciton fluid,
$e^2\delta C^{-1}=\partial\mu_{\rm ex}/\partial n_{\rm ex}$, and $d^*$
collapses from ${\approx}\,d$ to ${\sim}\,d^2/a_{\rm ex}$, set by the
dipole-dipole repulsion ($a_{\rm ex}$ is the interexciton spacing). The
quantum Hall counterpart of this collapse, $d^*\simeq0.6\,d^2/\ell_B$, was
conjectured as an experimental signature of exciton condensation in
graphene double layers in a strong magnetic field~\cite{Skinner2013QH};
the growth of the negative correction on approach to the transition, the
ladder channel nearing its pole, is its equilibrium precursor, and the
character of the anomaly at the transition itself (jump versus fluctuation
peak) reflects the order of the transition.

Beyond the specific results, the Lifshitz form of the interlayer
energy connects the capacitance problem to the asymptotic benchmarks of
dispersion-force theory~\cite{Bordag2009,Woods2016,Dobson2006}, and
Eq.~\eqref{eq:forcecap} relates the interlayer capacitance correction to
the Casimir pressure between the layers with a universal coefficient.

\begin{acknowledgments}
We thank Ilya Esterlis for carefully reading the manuscript, providing valuable comments, and drawing our attention to several useful references.  
We are grateful to Boris Shklovskii for his valuable feedback, for drawing our attention to Ref.~\cite{Skinner2013QH}, and for the discussions that followed.
This work was financially supported by National Science Foundation (NSF) Grant No. DMR-2452658 (D. Z. and A. L.) and H. I. Romnes Faculty Fellowship provided by the University of Wisconsin-Madison Office of the Vice Chancellor for Research and Graduate Education with funding from the Wisconsin Alumni Research Foundation (A. L.).  A. F. M gratefully acknowledges financial support from the Swiss National Science Foundation, Division II, under project 200021-227636. This work was performed in part during the workshop program "Emerging New Phases in Quantum Materials: The Disordered, the Strange and the Topological" at the Aspen Center for Physics, which is supported by National Science Foundation grant PHY-2210452. The authors acknowledge the use of the large language model Claude (Anthropic) in the preparation of the manuscript, including, in particular, numerical calculations, graphics, and symbolic verification of analytical results. The authors conceived the project and carried out the research.
\end{acknowledgments}

\appendix

\section{Polarization functions on the imaginary axis}
\label{app:lindhard}

\textit{2DEG.} For the spin-degenerate parabolic layer at $T=0$,
Eq.~\eqref{eq:lindhard} with $\mathrm{Re}\sqrt{\cdot}\ge0$; it reduces to
the Stern function at $\xi=0$, decays as $-nq^2/m\xi^2$ at high frequency,
and reproduces the hydrodynamic form
$\chi_0\to-\nu_0[1-\xi/\sqrt{\xi^2+v_F^2q^2}]$ at $q\ll\kF$. It was
verified against direct numerical evaluation of the polarization integral
at randomly chosen $(q,\xi)$.

\textit{Graphene.} The MLG and BLG bubbles are computed from
Eq.~\eqref{eq:bubble}. The MLG intrinsic part is exact,
$\Pi^{\rm MLG}_{\rm vac}=-(g/16)\,q^2/\hbar\sqrt{v^2q^2+\xi^2}$; the BLG
intrinsic part obeys the single-variable scaling
$\Pi^{\rm BLG}_{\rm vac}=-N_0\psi(2m\xi/\hbar q^2)$ with $\psi(0)=\ln4$
\cite{HwangKohn2008}, computed by quadrature from
Eq.~\eqref{eq:psi}; here $\xi$ is a frequency and $\hbar$ is restored. The
doped corrections are
evaluated with the principal-value-safe scheme of
Appendix~\ref{app:numerics} and validated as listed in
Sec.~\ref{sec:polarizability}.

\section{Plasmon zero-point integrals and universal constants}
\label{app:series}

With per-layer plasmons $\omega^2_{p,\ell}=Wq\,n_\ell^{\gamma}$ and
$a=e^{-qd}$, the coupled modes of the bilayer are
$\omega_\pm^2=\tfrac12(\omega_{p,1}^2+\omega_{p,2}^2)
\pm[\tfrac14(\omega_{p,1}^2-\omega_{p,2}^2)^2
+\omega_{p,1}^2\omega_{p,2}^2a^2]^{1/2}$, and
$E_{\rm int}/\A=(\hbar/4\pi)\int q\,\dd q\,
[\omega_++\omega_--\omega_{p,1}-\omega_{p,2}]$.

\textit{Energy.} For equal densities this yields Eq.~\eqref{eq:Easym}.
Expanding $\sqrt{1+a}+\sqrt{1-a}-2=2\sum_{k\ge1}\binom{1/2}{2k}a^{2k}$ and
using $\int_0^\infty u^{3/2}e^{-2ku}\dd u=\Gamma(5/2)(2k)^{-5/2}$,
\begin{equation}
J=\frac{3\sqrt\pi}{2}\sum_{k\ge1}\binom{1/2}{2k}(2k)^{-5/2}
=-0.0629780,
\end{equation}
whose terms decrease as $k^{-4}$; ten terms give four digits.

\textit{Curvature.} Setting $n_{1,2}=n(1\pm x)$, expanding to $O(x^2)$, and
using the stable identity
$(1+a)^{-1/2}-(1-a)^{-1/2}=-2a/[\sqrt{1-a^2}(\sqrt{1+a}+\sqrt{1-a})]$, the
curvature integrand is
\begin{align}
D_\gamma(a)&=\frac{\gamma(\gamma-1)}{2}
\left[\frac{1}{\sqrt{1+a}}+\frac{1}{\sqrt{1-a}}\right]
-\frac{\gamma(\gamma-2)}{2}
\nonumber\\
&\quad
-\frac{\gamma(\gamma-a^2)}{\sqrt{1-a^2}\,(\sqrt{1+a}+\sqrt{1-a})},
\label{eq:Dgamma}
\end{align}
and $\delta(1/C)_{\rm inter}=({\hbar\sqrt{Wn^\gamma}}/{4\pi e^2n^2})\,
K_\gamma d^{-5/2}$ with $K_\gamma=\int_0^\infty u^{3/2}
D_\gamma(e^{-u})\dd u$. For $\gamma=1$, Eq.~\eqref{eq:Dgamma} collapses to
Eq.~\eqref{eq:Dfun} and $K_1\equiv K=0.0477118$
[series $\Gamma(\tfrac52)(\tfrac3{16}2^{-5/2}+\tfrac{17}{256}4^{-5/2}
+\cdots)$]; for $\gamma=\tfrac12$ (Dirac), $K_{1/2}=0.0198002$.

\textit{Cross fraction.} Along the symmetric direction the curvature
integrand is $D_S(a)=-\tfrac14(\sqrt{1+a}+\sqrt{1-a}-2)$, and
$-2\partial_{n_1}\partial_{n_2}/[\partial^2_{n_1}+\partial^2_{n_2}
-2\partial_{n_1}\partial_{n_2}]=\tfrac12(1-\langle D_S\rangle/\langle
D_1\rangle)=0.3345$ with $\langle\cdot\rangle=\int u^{3/2}(\cdot)\dd u$,
and exactly $\tfrac13$ from the leading $a^2$ terms; this matches the
finite-difference decomposition of the full ring integral to all digits
computed.

\textit{Second-order tail.} With
$\int_0^\infty\dd\xi[1-\xi/\sqrt{\xi^2+v_F^2q^2}]^2=(2-\tfrac\pi2)v_Fq$,
Eq.~\eqref{eq:E2} gives $E_2/\A\to-(2-\tfrac\pi2)e^4\nu_0^2v_F/2d$, i.e.,
$G(x)\to(2-\pi/2)/2x$.

\section{Numerical methods and validation}
\label{app:numerics}

All integrals were evaluated with Gauss--Legendre quadratures; momentum
grids concentrate nodes at small $q$ with
$q_{\max}=16/2d+4\kF$, and frequencies are mapped by $\xi=\Omega_c\tan t$.
Capacitance curvatures are central second differences along
$(n+\delta n,n-\delta n)$ with steps $\delta n/n=0.02$--$0.05$ and
quadrature grids frozen across the stencil (removing the dominant noise
source); results are stable under step halving and grid doubling at the
$10^{-3}$ level or better.

\textit{2DEG.} Validation: the closed-form Lindhard function against
direct numerical integration ($\lesssim2\times10^{-3}$, limited by the
reference); $r_s$ independence of $E_2/N$ at fixed $\kF d$ (machine
precision); the analytic constants $c_0=2-\pi/2$ and Eq.~\eqref{eq:Dfun};
convergence onto Eqs.~\eqref{eq:Easym} and \eqref{eq:Kresult}
(Fig.~\ref{fig:2deg}). 

\textit{Graphene.} The bubbles \eqref{eq:bubble} are tabulated on
universal grids in $(q/\kF,\ \xi/v_Fq)$ ($200\times160$, cubic
interpolation) with three singular structures handled explicitly:
(a) the Fermi-surface crossing angle (principal value as $\xi\to0$),
integrated with log-spaced node pairs symmetric about the crossing so the
singular parts cancel pointwise; (b) the backscattering endpoint
$\theta=\pi$ (BLG only; the MLG coherence factor removes it); (c) a second,
radial principal-value structure at $k=q/2$, where rows on either side
diverge as $\pm1/(k-q/2)$, integrated with paired log-spaced nodes
symmetric about $q/2$. Omitting (b) or (c) produces $O(1)$ errors in the
BLG static limit, while the MLG case, which is less sensitive to
them, appears correct; the validation suite of Sec.~\ref{sec:polarizability} catches
both. Additional checks: the analytic intrinsic-MLG second-order energy
$E_2/\A=-\pi\alpha_{\rm g}^2g^2\hbar v/2048\,d^3$ is reproduced to
$10^{-4}$; the MLG second-order energy at fixed $\kF d$ is
$\kF$ independent in natural units to $10^{-6}$; the negative BLG
curvature at $\kF d=1.23$ was reproduced to three digits by a slow,
interpolation-free evaluation.

\bibliography{biblio-Capacitance}

\end{document}